\pdfoutput=1
\documentclass[10pt,a4paper]{article}
\usepackage[margin=23mm,headheight=14pt]{geometry}
\usepackage[T1]{fontenc}
\usepackage{lmodern,microtype}
\usepackage{amsmath,amssymb,booktabs,array,tabularx,longtable}
\usepackage{graphicx,xcolor,caption,float,placeins}
\newcommand{\rowsep}{\noalign{\vskip 2pt\hbox to\textwidth{\color{black!30}\leaders\hbox to 3pt{\hss\vrule height 0.4pt depth 0pt width 1.2pt\hss}\hfill}\vskip 2pt}}
\newcommand{\qhead}[1]{\par\smallskip\noindent\textit{#1}\enspace}
\usepackage{flafter}
\usepackage[numbers,sort&compress]{natbib}
\usepackage[colorlinks=true,linkcolor=blue!45!black,citecolor=blue!45!black,urlcolor=blue!45!black]{hyperref}
\usepackage{fancyhdr}
\newcommand{\AG}{\mathrm{AG}}
\newcommand{\AC}{\mathrm{OC}}
\newcommand{\W}{W_1}


\newcommand{\RepoLink}{\url{https://github.com/PuppyQ08/Raman_quality_measure}}
\title{\LARGE\bfseries Validating Spectral Quality Measures as Proxies\protect\\for Task Performance: A Controlled-Perturbation\protect\\Framework for Raman Spectroscopy}
\author{Xiuyi Qin\thanks{Email: terryqqy at gmail.com}\\[3pt]\small Independent Researcher\\\small\url{https://puppyq08.github.io/}}
\date{}
\hypersetup{pdftitle={Validating Spectral Quality Measures as Proxies for Task Performance: A Controlled-Perturbation Framework for Raman Spectroscopy},pdfauthor={Xiuyi Qin},pdfsubject={Raman spectroscopy benchmark and evaluation protocols}}

\begin{document}
\maketitle
\begin{abstract}
Raman preprocessing and enhancement methods are often assessed by comparing their output with a clean reference spectrum using measures such as mean squared error (MSE). Whether these spectral quality measures reflect downstream analytical performance is rarely tested. We propose a controlled-perturbation framework for this test. Five perturbation types (baseline distortion, independent noise, correlated noise, global wavenumber shift, and nonlinear axis warping) at eight strengths produce paired changes in a quality measure (metric harm) and in downstream performance (task harm). The alignment gap (AG) quantifies how much a single monotone metric-to-task relationship, fitted by isotonic regression, improves when each perturbation type receives its own relationship. Ordering concordance (OC), a clustered Kendall-type index, measures how often a measure ranks conditions of different perturbation types in the same order as their task harm. Cluster-bootstrap intervals and Holm-adjusted sign-flip tests compare twelve candidate measures with MSE. Three public datasets are used: bacterial classification by principal component analysis and logistic regression, sugar-mixture quantification by partial least squares regression, and mineral identification by cosine library matching. Models are fitted to unperturbed or to correspondingly perturbed training spectra. For bacterial classifiers fitted to unperturbed spectra, Wasserstein distance improved both statistics relative to MSE (Holm-adjusted $p<0.05$), also after axis perturbations were removed or spectra were compared on a common grid; refitting to perturbed spectra reversed this advantage. Peak-based measures and a structure-to-noise ratio improved both statistics only for refitted sugar calibrations, and no candidate did so for mineral identification. Open code supports testing new measures.
\end{abstract}

\section{Introduction}\label{sec:intro}
Raman spectroscopy probes the vibrational signatures of molecules and solids, making it a shared tool for studying composition, structure, and interactions across biology, chemistry, and materials science. Representative applications include label-free chemical imaging of biological samples \citep{freudiger2008}, the study of molecular adsorption at electrochemical interfaces \citep{fleischmann1974}, and the identification of graphene layers through their characteristic spectral features \citep{ferrari2006}. Across these applications, spectra serve as measurements to be processed, fingerprints to be compared, and observables against which theoretical descriptions can be tested. First-principles calculations and molecular dynamics provide routes to simulating Raman spectra, including approaches that use machine learning to predict polarizabilities and interatomic interactions \citep{lazzeri2003,raimbault2019,sommers2020}. Related vibrational theories describe anharmonic phonon dispersion and the thermodynamic effects of vibrations at finite temperature \citep{qin2020,qin2021,qin2023}. Quantifying spectral agreement is therefore a broad scientific need that extends from assessing preprocessing and enhancement to comparing calculated and experimental spectra.

Comparing a calculated Raman spectrum with an experimental one requires decisions about the wavenumber grid and intensity scaling. Calculations that produce discrete spectral lines also require a choice of line broadening: replacing each line with a peak of finite width to construct a continuous spectrum. With these choices specified, different measures capture different aspects of spectral agreement. We organize the thirteen measures evaluated here into four groups. \emph{Pointwise intensity errors} include mean squared error (MSE), root mean squared error (RMSE), mean absolute error (MAE), and normalized mean squared error (NMSE); they summarize differences between corresponding intensity samples. Such errors are widely used in Raman reconstruction assessment and analysis software \citep{barton2021,ramanspy2024}. \emph{Global shape measures}, spectral angle mapper (SAM) and Pearson correlation, compare the orientation of spectral vectors and their correlated intensity variation, respectively \citep{khan2012,mostafapour2026}. \emph{Physical-axis distribution distance} is represented by Wasserstein distance ($\W$), which measures how far normalized positive spectral intensity must move along the wavenumber axis to transform one distribution into the other. Related Wasserstein-based matching has been used to compare calculated and measured infrared spectra \citep{neymeyr2026}. \emph{Peak and structure measures} include peak precision, recall, F1, artifact ratio, missing ratio, and an auxiliary structure-to-noise ratio (S/N). The five peak outputs describe the preservation, addition, and loss of detected peaks; S/N compares spectral variation with an estimated noise level without requiring a reference spectrum. This group addresses the balance between noise suppression and peak preservation emphasized in Raman denoising research \citep{barton2021}. Exact definitions of the measures are given in the Supporting Information (SI), Section~S1.

In chemometrics, the accepted way to choose a preprocessing strategy is to judge it by the performance of the subsequent model. Reviews and systematic comparisons of spectral preprocessing judge candidate strategies by the resulting prediction or classification performance, and design-of-experiments procedures select preprocessing chains on the same basis \citep{rinnan2009,bocklitz2011,engel2013,gerretzen2015,liland2016,guo2021}. For Raman spectra, task-based studies connect preprocessing to concentration prediction, discrimination, and identification \citep{storey2019,martyna2020,carey2015,georgiev2024}. This task-based validation is reliable but specific: it requires labeled data, a chosen model, and a validation design \citep{westad2015}. Developers of smoothing, fluorescence correction, and learned enhancement methods \citep{savitzky1964,lieber2003,zhang2010,baek2015,gebrekidan2021,autoencoder2024} therefore also report spectral criteria that compare a processed spectrum with a clean reference spectrum, such as a high signal-to-noise measurement or a simulated spectrum before noise or background was added \citep{barton2021,mokari2026}, sometimes alongside downstream outcomes such as clustering of mineral maps \citep{noise2noise2026}. Raman-specific similarity metrics have been developed for library search \citep{khan2012,spcv2021}, and controlled studies have examined their sensitivity to noise, background, and wavenumber shifts \citep{mostafapour2026}. What remains largely untested is whether a spectral criterion is a valid \emph{proxy} for task performance. A weak, narrow band illustrates the problem: changing it may produce a small MSE yet impair classification if it distinguishes bacterial classes. A broad background distortion may produce a larger MSE with little effect on accuracy if the model tolerates it.

The answer may also depend on how the model was calibrated. Chemometric practice has two established responses to spectral variation that is absent from the calibration data. One tests the robustness of an existing model to that variation, as in studies of temperature effects and calibration transfer \citep{wulfert1998,feudale2002,wavelength2024,lora2025,umprecht2025}. The other includes the variation in the calibration set, as in robust calibration design and spectral data augmentation \citep{swierenga1999,bjerrum2017}. A defect that harms the first kind of model may be harmless to the second, so a quality measure that tracks one need not track the other.

We address these questions with a framework that treats a spectral quality measure as a proxy to be validated against task outcomes. Its contributions are as follows.
\begin{itemize}
\item A controlled-perturbation design in which every candidate measure is paired with identical downstream predictions, so that differences between measures are not confounded by changes in the model.
\item Two complementary statistics. The alignment gap (AG) uses isotonic regression to ask whether a measure needs a different interpretation for each type of spectral defect; ordering concordance (OC), a clustered Kendall-type index, asks whether the measure ranks defects of different types by their task harm.
\item Paired cluster-level inference against MSE, and a comparison between models fitted to unperturbed data (Fixed) and to perturbed data (Adapted).
\item A demonstration on bacterial classification \citep{ho2019}, sugar-mixture quantification \citep{sugars,georgiev2024}, and mineral identification \citep{lafuente2015}, with robustness controls for coordinate handling and perturbation set.
\end{itemize}
The complete comparison identifies different useful measures for fixed bacterial classification and condition-matched sugar quantification, with more limited evidence for mineral retrieval. The framework builds on open data and analysis tools \citep{ramanspy2024,sineesh2026,ramanbench2026}; the implementation, frozen experiment configurations, and all aggregate results are openly available at \RepoLink{}.

\section{Theory}\label{sec:theory}
\begin{figure}[!htbp]
\centering\includegraphics[width=\textwidth]{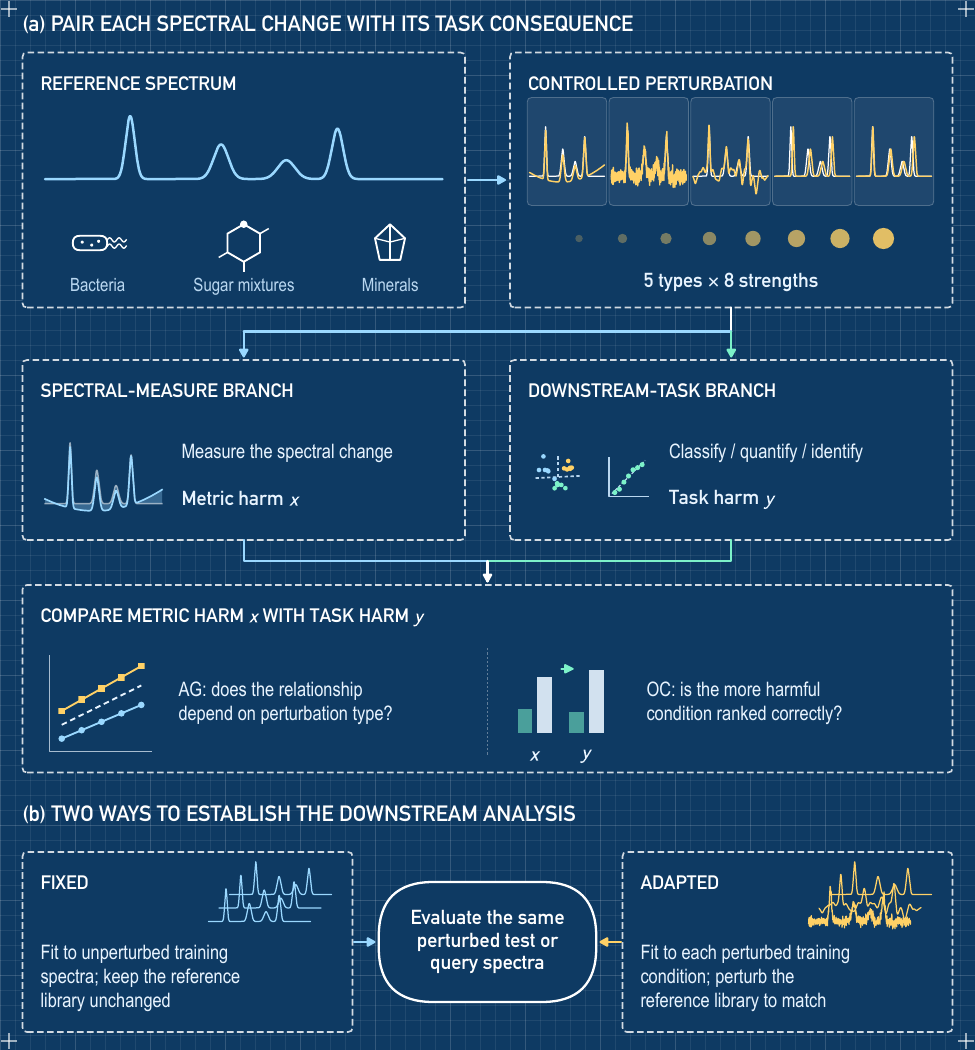}
\caption{The evaluation framework. (a) Each controlled perturbation yields a metric harm $x$, the change in a spectral quality measure, and a task harm $y$, the change in downstream loss. Every candidate measure is paired with the same task outcomes, so changing the measure does not change the predictions. (b) Fixed and Adapted analyses establish the downstream analysis from different training data or reference libraries and evaluate the same perturbed test or query spectra; all sample roles stay fixed. Pictograms and mini-plots are schematic and drawn from synthetic curves.}\label{fig:framework}
\end{figure}

Figure~\ref{fig:framework}a outlines the procedure. For each controlled perturbation, we record the change in a spectral quality measure and the change in downstream performance; AG and OC then compare the two across conditions. This section defines the quantities for a generic task. Section~\ref{sec:methods} specifies the perturbations, datasets, and models used here.

\subsection{Metric harm and task harm}\label{sec:harm}
A \emph{condition} is a pair $(p,\alpha)$ of perturbation type $p\in\{1,\ldots,P\}$ and strength $\alpha>0$; $\alpha=0$ denotes the unperturbed reference. Spectra are grouped into clusters $c=1,\ldots,C$, such as bacterial isolate classes. Clusters, rather than individual spectra, are resampled with replacement to assess how stable the estimates of AG, OC, and their differences between measures are, which yields their confidence intervals (Section~\ref{sec:inference}). Let $q$ be the cluster mean of a measure's per-spectrum values under a condition and $q_0$ its value for the unperturbed condition. We orient its change so that larger values indicate worse spectral quality:
\begin{equation}
x=\begin{cases}q-q_0,&\text{lower-is-better measure},\\q_0-q,&\text{higher-is-better measure}.\end{cases}
\label{eq:harm}
\end{equation}
This quantity is \emph{metric harm}. \emph{Task harm}, $y$, is unperturbed accuracy minus perturbed accuracy for classification and retrieval, or perturbed loss minus unperturbed loss for calibration. Negative values indicate improvement relative to the corresponding reference. Each cluster and condition contributes one $(x,y)$ observation. All candidate measures share the same $y$ values, so a comparison between measures isolates how they describe identical changes in task performance.

\subsection{Alignment gap}\label{sec:ag}
\qhead{AG: does the measure need a different interpretation for each perturbation?}
If similar metric harms accompany small task harms under noise but large task harms under a shift, a single relationship between the two quantities fits poorly (Figure~\ref{fig:agoc}a, right). We therefore fit task harm against metric harm in two ways. The \emph{pooled} fit is one nondecreasing curve $f$ for all observations, as if the measure meant the same thing under every perturbation; the \emph{separate} fits are one curve $f_p$ per perturbation type, which allow that meaning to differ. All clusters enter these isotonic fits together. AG is the reduction in squared residuals achieved by the separate fits, divided by the total sum of squares of task harm, $\mathrm{SST}=\sum_j(y_j-\bar y)^2$, so that it does not depend on the units of task harm:
\begin{equation}
\AG=\frac{\sum_j(y_j-f(x_j))^2-\sum_j(y_j-f_{p_j}(x_j))^2}{\mathrm{SST}},
\label{eq:ag}
\end{equation}
where $j$ indexes the $(x,y)$ observations and $p_j$ is the perturbation type of observation $j$. Equivalently, with $R^2=1-(\text{residual sum of squares})/\mathrm{SST}$ for each fit, AG is the gain $R^2_{\text{separate}}-R^2_{\text{pooled}}$ from allowing type-specific curves. It plays the role of a type-by-metric interaction in analysis of covariance, without assuming linearity. Monotonicity is the minimal assumption behind using a measure as a proxy: more metric harm should not predict less task harm. It also accommodates the saturating and threshold-like responses of bounded outcomes such as accuracy. The curves are least-squares isotonic regressions computed by pool-adjacent-violators, with tied metric harms receiving a common fitted value \citep{robertson1988}. Each type-specific fit can reproduce the pooled curve on its own observations, so $\AG\ge0$.

A smaller AG means that specifying the perturbation type adds less to the fitted relationship. This is an in-sample diagnostic: both fits can be poor even when their gap is small. A constant task harm makes AG undefined. Comparisons between monotone group relationships have an established statistical background \citep{durot2013}; here their reduction in residual error is the diagnostic of interest.

\subsection{Ordering concordance}\label{sec:oc}
\begin{figure}[!htbp]
\centering\includegraphics[width=\textwidth]{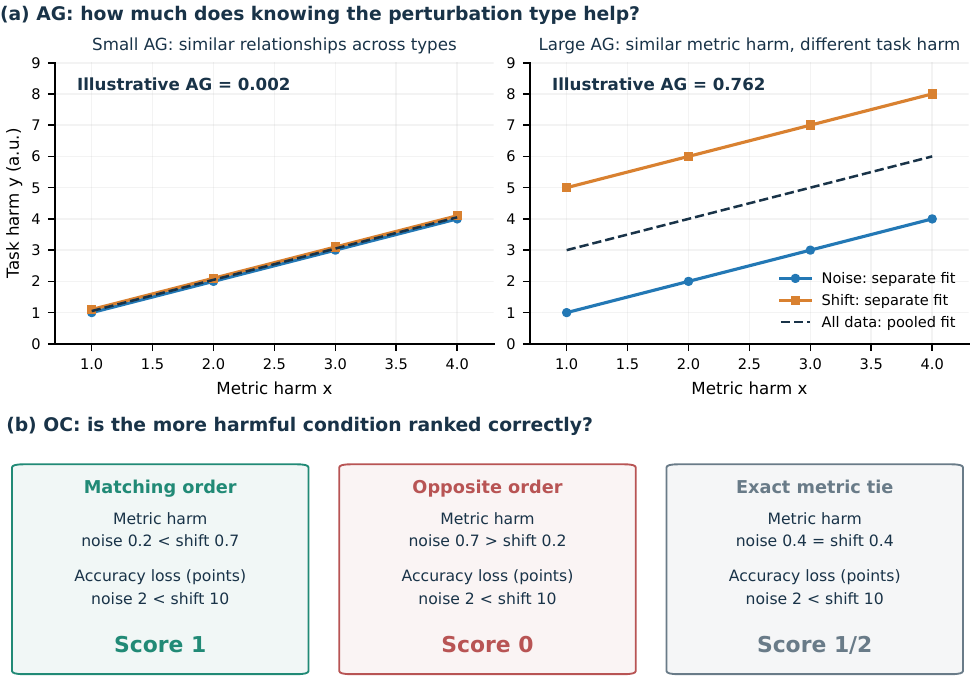}
\caption{AG and OC illustrated with synthetic values. (a) One pooled nondecreasing fit (dashed) is compared with a separate fit for each perturbation type (solid). Left: both types follow nearly the same relationship, so separate fits add little and AG is small. Right: equal metric harms accompany different task harms, so the pooled fit misses both types and AG is large. The experiments compare one pooled curve with five type-specific curves; a small AG can also occur when both fits are poor. (b) A pair of conditions from different perturbation types scores one when metric harm and task harm order it in the same way, zero when they disagree, and one-half for an exact tie in either quantity. OC averages these scores within each cluster and then across clusters.}\label{fig:agoc}
\end{figure}

\qhead{OC: does the measure identify the more harmful condition?}
Within each cluster, we compare conditions from different perturbation types (Figure~\ref{fig:agoc}b). A pair is \emph{concordant} (score one) when the condition with the larger metric harm also has the larger task harm, \emph{discordant} (score zero) when the two orderings are opposite, and \emph{tied} (score one-half) when either harm is exactly equal. With $\mathcal P_c$ the eligible pairs in cluster $c$, ordering concordance is
\begin{equation}
\AC=\frac{1}{C}\sum_{c=1}^C\frac{1}{|\mathcal P_c|}\sum_{(u,v)\in\mathcal P_c}\operatorname{score}(u,v).
\label{eq:oc}
\end{equation}
With $n_{\rm con}$, $n_{\rm dis}$, and $n_{\rm tie}$ the numbers of concordant, discordant, and tied pairs in cluster $c$, the inner average is $\AC_c=(n_{\rm con}+0.5\,n_{\rm tie})/|\mathcal P_c|$. It equals $(1+\tau_{a,c})/2$, where $\tau_{a,c}=(n_{\rm con}-n_{\rm dis})/|\mathcal P_c|$ is Kendall's rank correlation over the eligible pairs \citep{kendall1938}. For example, if 70 of 100 pairs are concordant, 20 are discordant, and 10 are tied,
\begin{equation}
\tau_{a,c}=\frac{70-20}{100}=0.50,\qquad \AC_c=\frac{70+0.5\times10}{100}=0.75=\frac{1+0.50}{2}.
\label{eq:ocexample}
\end{equation}
A measure that orders every pair correctly reaches $\AC_c=1$, and one unrelated to task harm is expected to score about 0.5. OC is thus a concordance probability of the kind used by Harrell's C-index to evaluate ranking \citep{harrell1982}; pairwise agreement is also used in metric meta-evaluation \citep{deutsch2023}. The restriction to pairs of different perturbation types is the essential design choice. Within one type, both harms usually increase with strength, so those pairs would mainly reward any measure that increases with $\alpha$. Cross-type pairs pose the question a practitioner faces: which of two different spectral defects is more harmful for the analysis? The five-type analysis used here contains $\binom52\times8^2=640$ pairs per cluster. OC evaluates ordering, rather than the magnitude of task harm or classification accuracy.

OC is unchanged by a strictly increasing transformation applied to all metric harms within a cluster. AG pools clusters, so different transformations across clusters can change its fit. These statements apply to the aggregated $x$ values: averaging per-spectrum RMSE or NMSE does not generally produce a monotone transformation of the cluster-mean MSE. This distinction prevents interpreting every small OC difference as an exact algebraic equivalence.

\subsection{Contrasts and inference}\label{sec:inference}
For measure $m$, define $\Delta\AG_m=\AG_{\rm MSE}-\AG_m$ and $\Delta\AC_m=\AC_m-\AC_{\rm MSE}$. Positive contrasts favor the candidate. MSE serves as the comparator because it is the most common default for comparing processed and reference spectra; the SI reports AG and OC for every measure, so any other pair can be compared. We report improvement on both summaries only when both contrasts are positive and both adjusted tests pass 0.05.

Each contrast is estimated from a limited number of clusters, so we ask two questions of it: how precisely is it estimated, and does it differ from zero? For precision, we resample clusters with replacement 2,000 times, keeping all conditions of a sampled cluster together and using the same draws for the candidate and MSE, and recompute the contrast with refitted isotonic curves. The 2.5th and 97.5th percentiles of these values give a pointwise 95\% interval \citep{field2007}. For the test, each cluster contributes a paired difference to the contrast. If the candidate and MSE agreed equally well with task harm, each contribution would be equally likely to be positive or negative. We therefore flip their signs at random 100,000 times and take the proportion of flipped totals at least as extreme as the observed one as a two-sided $p$-value. The sign-flip test holds the observed fitted curves fixed, whereas bootstrap draws refit them. Each task/protocol involves 24 such tests (twelve candidates, two summaries), and some would pass 0.05 by chance alone; Holm correction \citep{holm1979} keeps the probability of any false positive within this family at or below 0.05. Paired protocol effects (Section~\ref{sec:protocols}) form a separate family, and SI Table~S1 specifies all family sizes. Significance follows the adjusted tests; the displayed intervals are pointwise, that is, not adjusted for multiple comparisons.

\section{Materials and methods}\label{sec:methods}
\subsection{Datasets and downstream analyses}\label{sec:data}
The three datasets provide different definitions of successful analysis: a correct bacterial label, accurate mixture concentrations, and a correct mineral identification (Table~\ref{tab:data}). All metric comparisons within a task use the same data splits and downstream predictions.

Bacteria-ID supplies separate reference, finetuning, and test measurements \citep{ho2019}. We use its released polynomial-corrected, min--max-normalized spectra. Full-data and nested 5-, 10-, and 20-shot training sets are evaluated on 3,000 test spectra. Correctness is averaged within each of 30 isolate classes and across five fitting seeds. The isolate class is the statistical unit; the task measures isolate classification rather than patient-level diagnosis.

The sugar resource contains repeated low-SNR measurements of sucrose, fructose, maltose, and glucose mixtures \citep{sugars,georgiev2024}. Each of 240 physical wells contributes 32 acquisitions. The four concentration targets are nominal preparation values, with levels of 0, 0.08, 0.20, and 0.32 mol/L. Repeated acquisitions from a well stay together in five-fold evaluation. The loss averages squared concentration error over acquisitions and analytes and divides it by $(0.32\ \mathrm{mol/L})^2$.

The RRUFF cohort contains 3,770 spectra from 681 mineral classes \citep{lafuente2015}. Five fixed library/query splits separate related records using 1,934 grouping units. They yield 6,621 query occurrences from 3,012 unique query spectra. Query correctness is pooled within mineral class across splits, and mineral classes receive equal weight.

\begin{table}[!htbp]
\caption{Datasets, fitting sets, and downstream analyses. The final column gives the unit resampled for uncertainty. Test and validation roles remain separate from fitting.}\label{tab:data}\small
\begin{tabularx}{\textwidth}{@{}>{\raggedright\arraybackslash}p{27mm}>{\raggedright\arraybackslash}X>{\raggedright\arraybackslash}p{32mm}>{\raggedright\arraybackslash}p{20mm}@{}}
\toprule Task & Fitting and evaluation data & Analysis & Statistical unit\\\midrule
Bacteria, full & 60,000 reference and 2,700 finetuning training spectra; 300 validation; 3,000 test & 20-component PCA; logistic regression & 30 isolate classes\\
Bacteria, few shot & 5/10/20 training spectra per class; 300 validation; 3,000 test; five fitting seeds & Same PCA and logistic procedure & 30 classes per endpoint\\
Sugar mixtures & 240 wells; 32 acquisitions/well; five folds, each with 144 training, 48 validation, 48 test wells & Multi-output PLS regression & 240 wells\\
Minerals & 3,770 spectra; 681 minerals; five library/query splits & L2-normalized cosine nearest match & 681 mineral classes\\
\bottomrule\end{tabularx}
\end{table}

Bacteria use 20-component PCA without whitening followed by L2-regularized logistic regression, with $C\in\{0.01,0.1,1,10\}$ selected by validation accuracy. Sugar uses PLS regression \citep{wold2001} with 2, 4, 8, 16, or 32 components, selected by validation macro normalized RMSE. The selected fit is evaluated without adding validation observations to training. Mineral matching uses cosine similarity after L2 normalization. These are standard, transparent chemometric analyses, chosen so that the spectral measures rather than model complexity are the object of study. The framework requires only paired task outcomes and applies unchanged to other models.

\subsection{Controlled perturbations}\label{sec:perturb}
\begin{figure}[!htbp]
\centering\includegraphics[width=\textwidth]{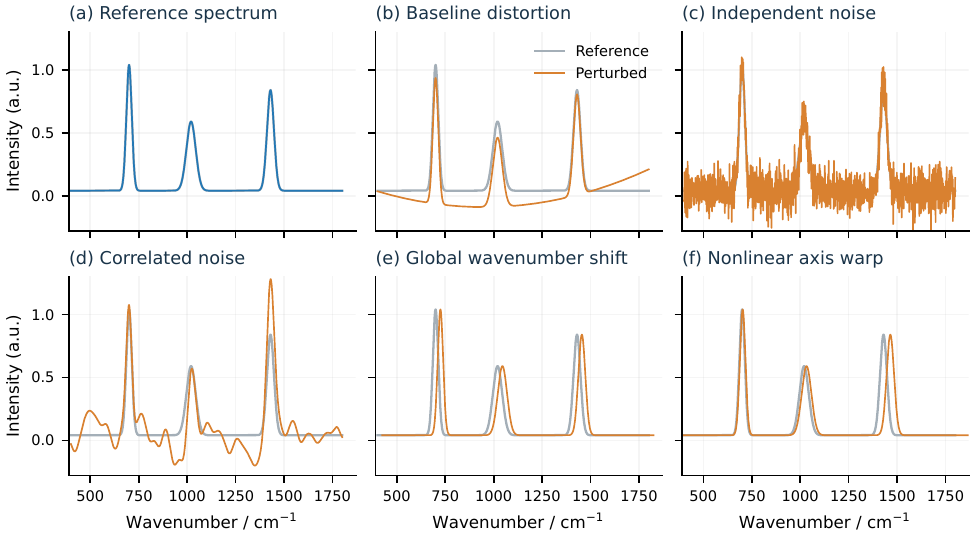}
\caption{Synthetic illustrations of the five perturbation types in Table~\ref{tab:perturb}. Gray curves show the reference spectrum and orange curves the perturbed spectrum. The shift and warp are enlarged for visibility; in the experiments the largest coordinate displacement is 3.2~cm$^{-1}$. These are teaching examples, not measured spectra.}\label{fig:perturb}
\end{figure}

We change one source of spectral variation at a time. Each spectrum receives baseline distortion, independent Gaussian noise, correlated noise, a global shift, and a quadratic axis warp (Table~\ref{tab:perturb}; Figure~\ref{fig:perturb}). For a spectrum with intensities $s=(s_1,\ldots,s_n)$ at wavenumbers $\nu=(\nu_1,\ldots,\nu_n)$, let $r_s=(n^{-1}\sum_i s_i^2)^{1/2}$ be its root-mean-square (RMS) intensity. The three intensity perturbations add a random shape vector of the same length to $s$, scaled by the strength $\alpha$ and by $r_s$: a smooth background $b$ for baseline distortion, a vector $z$ of independent standard-normal values for independent noise, and a smoothed noise vector $\widetilde z$ for correlated noise. Scaling by $r_s$ expresses $\alpha$ as a fraction of the spectrum's own RMS intensity. The two axis perturbations instead move the wavenumbers $\nu$ and leave the intensities unchanged. Each operator uses eight strengths, $\alpha\in\{0.05,0.10,0.20,0.30,0.40,0.50,0.65,0.80\}$. A fixed realization is scaled across strengths for a given spectrum and operator. The unperturbed spectrum, $\alpha=0$, supplies the reference. The forty positive conditions enter the metric comparison.

The perturbations are not intended to reproduce the output of any particular preprocessing method. They isolate single defect types so that the dependence of a measure on defect type can be identified. The strengths range from a component with 5\% of the spectrum's RMS intensity to one with 80\%. The largest coordinate displacement, 3.2~cm$^{-1}$, is less than two sampling intervals of the 2~cm$^{-1}$ mineral grid.

Because the baseline and correlated-noise shapes $b$ and $\widetilde z$ have unit RMS, both give
\begin{equation}
\mathrm{MSE}=\alpha^2r_s^2,\qquad \mathrm{NMSE}=\alpha^2,\qquad \mathrm{RMSE}=\alpha r_s
\label{eq:energy}
\end{equation}
on the native intensity arrays. Independent noise has the same MSE and NMSE \emph{in expectation}; its finite realization is not normalized to unit RMS. Thus these perturbations compare different shapes at equal expected squared-error energy. MSE records their magnitude but cannot distinguish the baseline and correlated-noise shapes at a given strength. Finite independent-noise realizations and subsequent averaging can affect their ordering. SI Section~S2 gives the exact relationships.

\begin{table}[!htbp]
\caption{Five perturbation types. Here $s$ and $s'$ are the reference and perturbed intensities, $r_s$ is the RMS intensity of $s$, $\alpha$ is the strength, and coordinates $\nu$ are in cm$^{-1}$. The shape and random state are fixed for each spectrum/operator pair and reused across strengths.}\label{tab:perturb}\small
\begin{tabularx}{\textwidth}{@{}>{\raggedright\arraybackslash}p{28mm}X@{}}
\toprule Perturbation & Definition\\\midrule
Baseline distortion & $s'=s+\alpha r_s b$, where $b$ is a smooth background such as residual fluorescence or instrument drift: a randomly weighted Legendre expansion of orders 2 through a sampled degree in $\{2,3,4\}$, centered and normalized to unit RMS.\\
Independent noise & $s'=s+\alpha r_s z$, where $z$ has independent standard-normal entries. The realized noise is not rescaled.\\
Correlated noise & $s'=s+\alpha r_s\widetilde z$, where $\widetilde z$ is white noise filtered by a row-normalized Gaussian kernel with width 20 cm$^{-1}$, then centered and normalized to unit RMS.\\
Shift & $\nu'_i=\nu_i+4\alpha$, with $s'_i=s_i$.\\
Warp & $\nu'_i=\nu_i+4\alpha[(\nu_i-\nu_{\min})/(\nu_{\max}-\nu_{\min})]^2$, with $s'_i=s_i$.\\
\bottomrule\end{tabularx}
\end{table}

\subsection{Fixed and Adapted analyses}\label{sec:protocols}
Training spectra establish the bacterial classification rule or the sugar concentration calibration. Validation spectra select its settings; test spectra measure performance. Mineral identification instead compares a query with known reference spectra. Figure~\ref{fig:framework}b contrasts the two analyses, which correspond to the two chemometric strategies described in Section~\ref{sec:intro}.

In the \textbf{Fixed} analysis, PCA/logistic or PLS parameters are fitted and selected using unperturbed training and validation spectra, then held fixed while the test spectra are perturbed. The mineral library remains unchanged. This analysis tests a model's robustness to variation absent from its calibration data \citep{wulfert1998,feudale2002}. In the \textbf{Adapted} analysis, training and validation spectra receive the same perturbation type and strength as the test spectra, and a separate fit is selected for each condition. Both PCA and logistic parameters are refitted for bacteria. The mineral library receives the same nominal perturbation as its queries, using distinct record-specific random draws. This analysis includes the variation in the calibration data, the strategy behind robust calibration designs \citep{swierenga1999} and spectral data augmentation, in which perturbed copies of training spectra are added to the calibration set \citep{bjerrum2017}. Both protocols evaluate identical perturbed test or query spectra and maintain the same sample roles. The controlled comparison is not a cross-instrument transfer benchmark. Full-data bacteria has only the Fixed analysis; the three few-shot endpoints, sugar, and minerals have both, giving eleven task/protocol combinations.

The two analyses are compared in two ways. For each perturbation type $p$, the protocol effect
\begin{equation}
G_p=\operatorname*{mean}_{c,\alpha}\left(y^{\rm Fixed}_{cp\alpha}-y^{\rm Adapted}_{cp\alpha}\right)
\label{eq:gp}
\end{equation}
is the mean change in task harm over clusters $c$ and strengths $\alpha$; positive values indicate less harm in the Adapted analysis. For each measure, the metric interaction
\begin{equation}
I=\Delta^{\rm Adapted}-\Delta^{\rm Fixed}
\label{eq:interaction}
\end{equation}
is the change in its MSE-relative advantage, where $\Delta$ is $\Delta\AG$ or $\Delta\AC$ (Section~\ref{sec:inference}) computed under each analysis. It is calculated separately for AG and OC using shared cluster draws. A negative $I$ means that the advantage over MSE shrinks or reverses when the analysis is fitted to perturbed data.

\subsection{Spectral representation and quality measures}\label{sec:representation}
Metrics are calculated either on each spectrum's own coordinates and intensities (\emph{native}) or after both spectra are interpolated onto the task grid (\emph{common grid}). The downstream analysis always receives the task-grid spectrum. Bacteria use 997 source-grid coordinates, sugar 1,999 coordinates, and minerals 799 coordinates from 204 to 1,800 cm$^{-1}$. Each fixed support covers all transformations without extrapolation. Coordinate shifts preserve the native intensity vector, so index-aligned intensity errors remain zero for these operators. On the common grid, the same shifts change the sampled intensities and become visible to MSE.

Wasserstein distance weights positive intensities by trapezoidal node widths and normalizes their total mass to one. Peak measures use a continuous-wavelet detector with widths 1, 2, 4, and 8 cm$^{-1}$, minimum SNR 2, noise percentile 10, and a 2 cm$^{-1}$ matching tolerance. For a single pair of spectra, the hit quality index used in library search is a monotone function of cosine similarity, and hence of SAM, so its orderings before averaging are represented by SAM. Metrics are computed per spectrum before averaging within the statistical unit of Table~\ref{tab:data}, hereafter a cluster; fitting seeds and split occurrences are aggregated before inference. Definitions and numerical settings are in SI Section~S1.

\subsection{Robustness controls}\label{sec:controls}
We call the analysis with native metrics and all five perturbation types the \emph{main analysis}. After examining its results, we added controls to test whether the W1 advantage depended on axis perturbations or coordinate handling. Two factors define four analyses: native versus common-grid metrics, and all five perturbation types versus the three intensity types. All thirteen measures are recomputed in the common-grid representation, with their own unperturbed reference values. The downstream predictions stay fixed. Removing the two axis types (shift and warp) requires new AG fits and reduces OC to 192 pairs per cluster. The three control analyses use separate Holm families defined before their execution: 66 W1--MSE tests and 726 other-candidate tests. They are post hoc robustness analyses. SI Sections~S4--S5 provide full contrasts, decompositions, and rules for undefined results.

\subsection{Software}\label{sec:software}
All computations used Python 3.13.11 with NumPy 2.5.2, SciPy 1.18.0, scikit-learn 1.9.0, and Matplotlib 3.11.1 \citep{scipy2020,sklearn2011}. Random states, grids, and model settings are recorded in frozen configuration files in the accompanying repository (SI Section~S6).

\section{Results}\label{sec:results}
\subsection{The complete comparison across tasks}\label{sec:res-overview}
\begin{figure}[!htbp]
\centering\includegraphics[width=\textwidth]{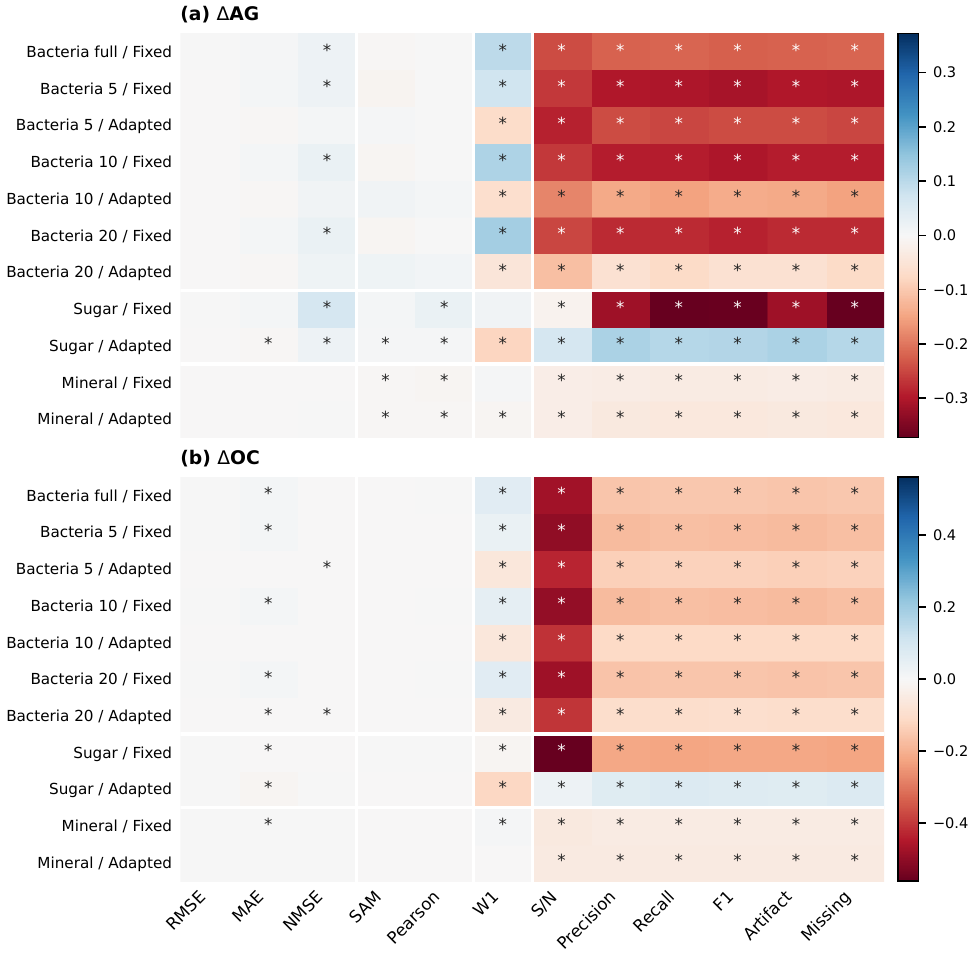}
\caption{Results of the main analysis (native metrics, all five perturbation types in Table~\ref{tab:perturb}) for every candidate relative to MSE. Each cell summarizes one task/protocol (row) and one candidate measure (column) over all perturbation types and strengths. Positive contrasts (blue) favor the candidate; negative contrasts (red) favor MSE. The four column groups are intensity errors, global shape, physical-axis distance, and peak/structure measures; MSE is the reference and has no column. The two panels have separate scales. Asterisks mark two-sided Holm-adjusted $p<0.05$ among the 24 tests for each task/protocol. Improvement on both summaries requires positive significant entries in both panels.}\label{fig:alignment}
\end{figure}

Figure~\ref{fig:alignment} shows the main analysis for all twelve candidates in the eleven task/protocol combinations; SI Section~S3 gives all 143 metric/task/protocol rows. Two patterns stand out. W1 improves both summaries for all four Fixed bacterial endpoints. The five peak outputs (precision, recall, F1, artifact ratio, and missing ratio; hereafter peak stability) and S/N improve both only for Adapted sugar quantification and are worse on both in the other ten combinations. Other candidates have at most one significant improvement in a given combination. Sections~\ref{sec:res-w1fixed} and~\ref{sec:res-w1adapted} report the W1 results under the Fixed and Adapted analyses, Section~\ref{sec:res-peak} the peak-stability and S/N results, and Section~\ref{sec:res-other} the remaining measures and mineral retrieval.

\subsection{Wasserstein distance with fixed bacterial classifiers}\label{sec:res-w1fixed}
W1 tracks task harm better than MSE on both summaries at every training-set size. At 10 shots, it lowers AG from 0.16 to 0.046 and raises OC from 0.78 to 0.83; both contrasts pass the adjusted tests. The advantage persists under the controls (Figure~\ref{fig:w1}a,b): all four Fixed bacterial endpoints improve on both summaries with the axis perturbations removed, with common-grid metrics, and with both changes, and all 24 corresponding tests pass the 66-test robustness adjustment. At 10 shots, removing the axis perturbations increases the favorable contrasts to 0.17 for AG and 0.12 for OC, and applying both controls gives 0.17 (95\% CI 0.13--0.22) and 0.15. In the main analysis, the two axis types contribute only 0.65\% of MSE's AG in this setting, and non-axis pairs account for 71\% of the W1--MSE OC gain. Full contrasts and decompositions appear in SI Sections~S4--S5.

\subsection{Fitting to perturbed spectra changes the W1 comparison}\label{sec:res-w1adapted}
\begin{figure}[!htbp]
\centering\includegraphics[width=\textwidth]{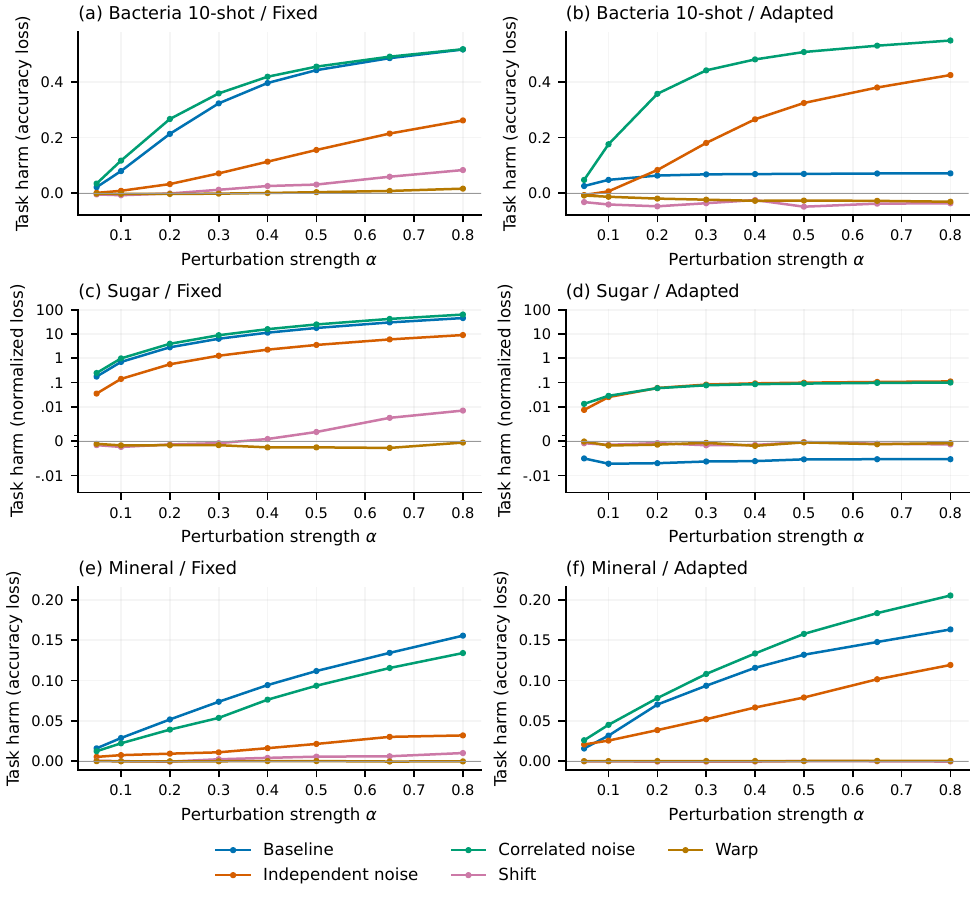}
\caption{Task harm under Fixed and Adapted analyses. Points are cluster-mean values at eight perturbation strengths. Bacterial and mineral task harm is accuracy loss; sugar task harm is normalized squared-loss increase, shown on a symmetric-log scale with a linear region around zero. Within each task, the two panels share their vertical range. Uncertainty for the mean protocol effects is reported in SI Table~S3.}\label{fig:responses}
\end{figure}

The Adapted analysis changes which perturbations cause the greatest task harm (Figure~\ref{fig:responses}). For 10-shot bacteria, fitting to perturbed data reduces baseline harm by $G_{\rm baseline}=0.25$ (95\% CI 0.20--0.29) but increases independent- and correlated-noise harm, with $G=-0.10$ and $-0.054$. All three adjusted tests pass 0.05. Sugar behaves differently: the corresponding effects are positive for all three intensity perturbations, with reductions of 14, 2.8, and 20 in normalized squared-loss units. Perturbing the mineral library increases harm from both noise types, by about 0.05 in accuracy units. SI Table~S3 lists all protocol effects.

\begin{figure}[!htbp]
\centering\includegraphics[width=\textwidth]{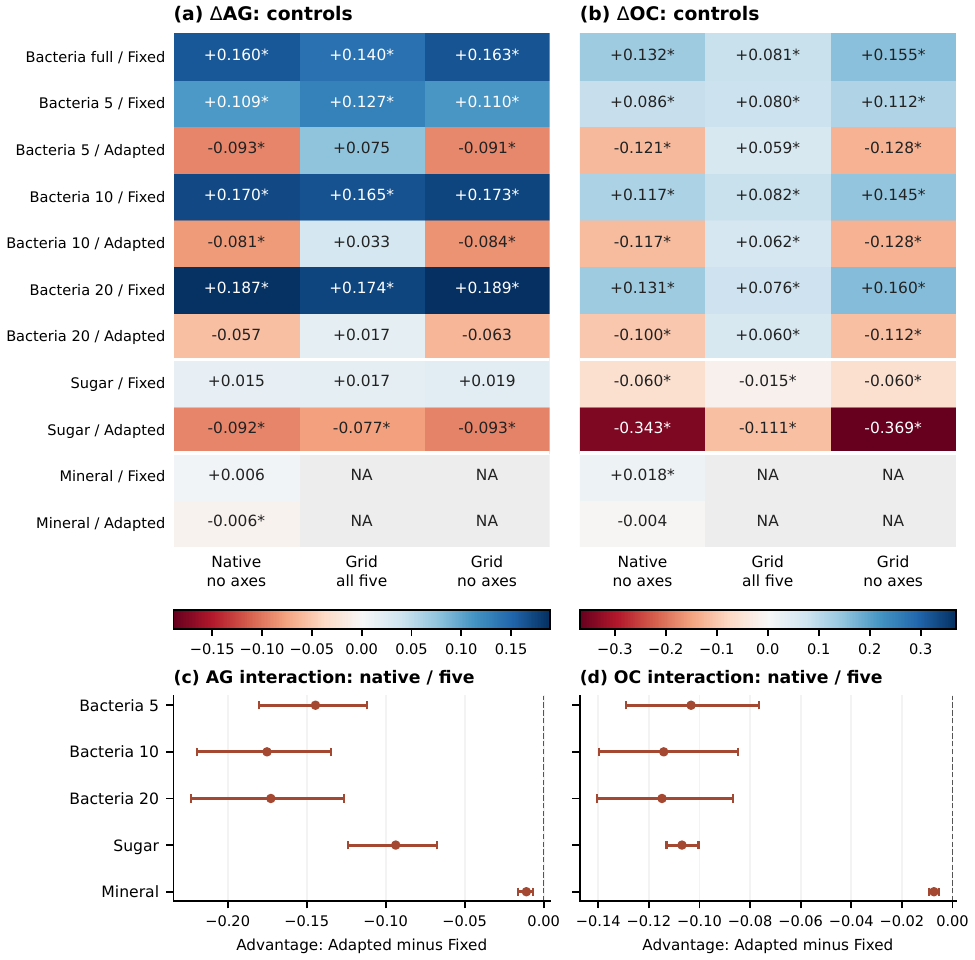}
\caption{W1 comparisons under controls and protocol changes. (a,b) Three robustness analyses; positive contrasts favor W1. Asterisks mark $p_H<0.05$ among the 66 W1 robustness tests. No axes retains baseline distortion and both noise types; grid means common-grid metrics. NA denotes unavailable mineral W1 results: four records give zero positive mass in 19 perturbed evaluations, making four analysis rows unavailable. (c,d) Changes in W1's MSE-relative advantage from Fixed to Adapted in the main analysis. Intervals are pointwise 95\% cluster-bootstrap intervals; all ten interactions pass their endpoint-specific 29-test adjustment. Complete estimates are in SI Tables~S4--S5 and accompanying CSVs.}\label{fig:w1}
\end{figure}

W1 loses its bacterial advantage under Adapted fitting in the main analysis. At 10 shots its AG and OC contrasts become $-0.062$ and $-0.065$, both significant. The same directions occur at 5 and 20 shots. W1's advantage decreases from Fixed to Adapted on both summaries in all five paired endpoints; all ten interaction tests pass their endpoint-specific adjustment (Figure~\ref{fig:w1}c,d).

The controls refine this reversal. With the axis perturbations removed, W1 has negative AG and OC contrasts at every Adapted shot count in both representations, although the 20-shot AG contrasts do not pass the robustness adjustment. With a common grid and all five perturbation types, OC instead favors W1 at all three shot counts, while none of their AG tests passes. At 10 shots, applying both controls leaves interactions of $-0.26$ (95\% CI $-0.35$ to $-0.20$) for AG and $-0.27$ for OC. The reversal persists without axis perturbations, but its form depends on the representation and perturbation set.

In sugar quantification, W1 does not gain an advantage under either analysis. With a common grid and no axis perturbations, Adapted sugar has $\Delta\AG=-0.093$ and $\Delta\AC=-0.37$, both significant; Fixed sugar has an adverse OC contrast and an unresolved AG contrast under each control.

\subsection{Peak stability and S/N in sugar quantification}\label{sec:res-peak}
Peak stability and S/N improve both summaries only for Adapted sugar quantification. Peak F1 reduces AG by 0.11 (95\% CI 0.070--0.15) and raises OC by 0.071 relative to MSE, and S/N improves the same summaries by 0.062 and 0.029. All four adjusted tests pass 0.05. Each of the five peak outputs and S/N is worse on both summaries in Fixed sugar and in every bacterial and mineral combination.

\subsection{Intensity measures, global shape, and mineral retrieval}\label{sec:res-other}
The intensity measures give smaller, partial gains. MAE improves OC in Fixed bacteria and Fixed mineral retrieval, and NMSE improves AG in Fixed bacteria and both sugar protocols; neither improves both summaries in these settings. RMSE has no significant favorable contrast. In 10-shot Fixed bacteria, MAE's OC gain includes independent-versus-correlated-noise pairs with no metric ties in either measure (SI Section~S2). Pearson improves AG in both sugar protocols, and SAM does so in Adapted sugar; their OC improvements are not significant.

Mineral retrieval supplies little evidence for replacing MSE. With the Fixed library, W1 improves OC by 0.006, but its AG contrast is not significant after adjustment; MAE also improves OC alone. With the Adapted library, W1 has a worse AG and an unresolved OC difference. In the main analysis, no candidate improves both summaries under either library protocol. Removing the axis perturbations leaves a partial W1 gain for the Fixed library. Common-grid W1 is undefined for four records; the affected comparisons are reported as unavailable rather than estimated on a selectively reduced cohort, and the other twelve measures remain evaluable.

\FloatBarrier
\section{Discussion}\label{sec:discussion}
\subsection{Principal findings and practical guidance}
Four findings stand out. Together they describe when a spectral quality measure can stand in for task performance.

\textbf{No measure replaces MSE universally.} Each candidate that improves both summaries does so only in specific settings: W1 for classifiers fitted to unperturbed bacterial spectra, and peak stability and S/N for calibrations refitted to perturbed sugar spectra; mineral retrieval supports no replacement (Figure~\ref{fig:alignment}). The appropriate measure therefore depends on the analysis it is meant to represent, and Table~\ref{tab:guidance} summarizes the evidence for each analysis type. The five peak outputs are computed from the same detected peaks, so their joint result is one peak-stability profile rather than five independent confirmations. Because the PLS predictions are identical across measures, this result concerns how the measures assess concentration errors, not an improvement of PLS by a peak score. These recommendations concern the stated tasks and procedures.

\begin{table}[!htbp]
\caption{Practical implications of the main analysis. The Fixed bacterial W1 result also passes the axis-removal and common-grid controls. The remaining rows describe the main analysis; their scope is not extended automatically to other representations or analyses.}\label{tab:guidance}\small
\begin{tabularx}{\textwidth}{@{}>{\raggedright\arraybackslash}p{46mm}X@{}}
\toprule Intended analysis & Evidence from this study\\\midrule
Bacterial classifier fitted on unperturbed spectra & Report W1 alongside MSE: both summaries improve, including under the robustness controls.\\\rowsep
Few-shot classifier fitted to each perturbed condition & No candidate improves both summaries. W1, S/N, and peak outputs are worse on both in the main analysis.\\\rowsep
PLS calibration fitted on unperturbed spectra & No candidate improves both summaries; lower AG for some measures does not accompany better OC.\\\rowsep
PLS calibration fitted to each perturbed condition & Peak stability and S/N improve both summaries; W1 and MAE are worse on both.\\\rowsep
Cosine library identification & No candidate improves both summaries under either protocol; Fixed-library W1 gives a small OC improvement.\\
\bottomrule\end{tabularx}
\end{table}

\textbf{The W1 advantage is not an artifact of coordinate handling.} In the native representation, a shift or warp leaves the intensity vector unchanged, so MSE registers no change for these perturbations while W1 does. The Fixed bacterial advantage nevertheless persists when the axis perturbations are removed and when all measures are compared on a common grid, and non-axis pairs provide most of the OC gain (71\% at 10 shots; Section~\ref{sec:res-w1fixed}). The advantage therefore extends beyond detecting displaced coordinates. It is also more specific than a general advantage of transport distance, because sugar and mineral analyses show different profiles. A measure's physical interpretation is useful for understanding what it responds to; the task comparison determines whether those responses are informative for the chosen analysis.

\textbf{The calibration strategy can reverse a measure's advantage.} Refitting the bacterial classifier to perturbed spectra lowered baseline harm but raised noise harm (Figure~\ref{fig:responses}). W1's contrasts with MSE became negative at every few-shot size, and its advantage decreased from Fixed to Adapted in all five paired endpoints (Figure~\ref{fig:w1}c,d). A quality measure that tracks the harm to a model calibrated on clean spectra is suited to judging whether a processing step protects an existing model. After the model has been recalibrated on processed or augmented data, the same measure can mislead. A score validated under one calibration strategy therefore needs revalidation after the strategy changes, much as a calibration model needs revalidation after transfer \citep{feudale2002}.

\textbf{AG and OC capture different aspects of agreement, and related measures are not interchangeable.} Several measures improve one summary but not the other: NMSE lowers AG in Fixed bacteria and both sugar protocols without better OC there, and MAE raises OC in Fixed bacteria and Fixed mineral retrieval without lower AG (Section~\ref{sec:res-other}). A measure can therefore follow one relationship across perturbation types yet misorder conditions, or the reverse, and both summaries are needed. MSE, RMSE, and NMSE are algebraically related for each spectrum, yet they give different results because averaging within clusters does not preserve those relationships (Section~\ref{sec:oc}). MAE weights the same residuals differently, and its OC gain includes pairs with no metric ties, so breaking ties with the baseline alone does not explain it (SI Section~S2).

\subsection{Relative sensitivity as an interpretation}
The pattern is consistent with a simple interpretation: agreement improves when a measure distinguishes the perturbations that matter most to the downstream procedure. Refitting can change those priorities, as the protocol effects in Section~\ref{sec:res-w1adapted} show, so a fixed ordering of spectral changes may agree with one procedure and disagree with another.

This interpretation does not identify a single mechanism behind the matrix. PCA projection, PLS fitting, positive-mass normalization in W1, and the peak detector can each change sensitivity in different ways. A smooth baseline can increase the numerator of S/N while having a smaller effect on its first-difference noise estimate, but the sign also depends on the spectrum and the baseline. Likewise, narrow-scale wavelet detection can suppress broad background variation without being exactly invariant to it. Establishing which component causes a particular reversal would require component-specific comparisons. The present evidence establishes the conditional metric profiles and their robustness, rather than that causal decomposition.

\subsection{Applying the framework to a new measure or task}
Reuse requires a defined downstream task, a coordinate convention, a fitting or library protocol, and common observations for all candidate measures. In practice the procedure has five steps.
\begin{enumerate}
\item Fix the downstream analysis, its data splits, and the cluster that serves as the statistical unit.
\item Apply the perturbation operators at a grid of strengths to the evaluation spectra, and, for an Adapted analysis, to the fitting data or library.
\item Record the task outcome for every cluster and condition once; these outcomes are shared by all measures.
\item Compute each measure's metric harm (Eq.~\ref{eq:harm}) on the chosen representation.
\item Estimate AG and OC with paired cluster-bootstrap intervals and adjusted tests against the comparator.
\end{enumerate}
The accompanying repository includes a data-free example that carries out these steps for all thirteen measures and can be adapted to a new measure, task, or dataset (SI Section~S6). The same design could be applied to calculated-versus-measured spectra if a labeled task, such as phase identification, supplies an outcome. Broadening, scaling, and grid choices would then be explicit factors in that evaluation. The present datasets test experimental spectra; they supply a procedure for that extension rather than direct evidence about simulation accuracy. The framework complements the data and software infrastructure of RamanSPy and RamanBench \citep{ramanspy2024,ramanbench2026} by evaluating the spectral measures themselves.

Spectral measures also serve as training losses, which the present framework does not evaluate. A Raman denoising network was trained with a custom loss designed to limit corruption of spectral peaks \citep{barton2021}. For infrared spectra predicted from molecular structure by message passing neural networks, a preliminary controlled comparison found that training with MSE was outperformed by training with the spectral information divergence, RMSE, or a target-weighted MSE, although MSE and RMSE are monotonically related \citep{mcgill2021}. That study judged the trained models with a separate similarity score, which uses Gaussian convolution to give partial credit for shifted peaks, and noted that this score was unsuitable as a training loss because it produced aphysical prediction artifacts. The same distinction applies here: AG and OC test measures as assessment criteria against task harm. Whether a measure that better tracks task harm also yields a better training loss requires a separate study.

\section{Conclusions}\label{sec:conclusions}
We provide a controlled-perturbation framework for evaluating Raman spectral quality measures against task harm using AG and OC. AG measures, through isotonic regression, whether a measure needs a different interpretation for each type of spectral defect; OC measures, as a clustered Kendall-type concordance, whether it ranks different defects by their task harm. Across three applications, W1 improves both summaries for Fixed bacterial classifiers, and this finding survives axis removal and common-grid comparison. Peak stability and S/N improve both summaries specifically for Adapted sugar quantification in the main analysis; mineral retrieval identifies no replacement that improves both. Refitting and spectral representation can change these outcomes, so the validity of a quality measure is a property of the measure together with the analysis and its calibration strategy. The complete metric matrix and robustness controls give concrete evidence for choosing measures in comparable analyses and a common procedure for evaluating new candidates.

\section*{Data and code availability}
The source datasets are Bacteria-ID \citep{ho2019}, RRUFF \citep{lafuente2015}, and the Raman sugar-mixture archive \citep{sugars}. Source discovery and loading checks used RamanBench v0.1.1 and \texttt{raman-data} v1.2.6 \citep{ramanbench2026}; sugar files were accessed through its cache of the source archive. Cohorts, grouping, splits, and downstream experiments follow the definitions in this article.

Code and aggregate results are available at \RepoLink{} under the MIT License. The repository contains the implementations of the thirteen measures, the perturbation operators, and the AG/OC statistics and inference; the executed experiment pipelines with their frozen configurations; the 143-row native summary, the 572-row robustness summary, and the other aggregate tables behind every figure and table; the code that rebuilds these displays; and a data-free example of the complete procedure. Third-party spectra are not redistributed and remain subject to their source terms; scripts download them from the official sources and verify each archive's size and SHA-256 digest. SI Section~S6 describes the repository layout, the correspondence between article names and code identifiers, the file behind each reported result, and how to use the code. The Supporting Information also gives definitions, inferential details, and full numerical tables.

\section*{Acknowledgments}
The author thanks the maintainers of Bacteria-ID, RRUFF, the Raman sugar-mixture resource, RamanBench and raman-data, RamanSPy, and the scientific Python ecosystem. The author also thanks the AI assistants Codex (OpenAI) and Claude (Anthropic) for help with revising the manuscript, writing the plotting code for the figures, testing and running the main code, and checking the data and results. The author reviewed these contributions and takes responsibility for the content of this article.

\FloatBarrier
\begingroup\setlength{\bibsep}{3pt}
\bibliographystyle{unsrtnat}
\bibliography{references}
\endgroup
\end{document}